\documentclass[fleqn]{SCGE-v2}
\usepackage{listings}
\usepackage{multirow}
\usepackage{booktabs}
\usepackage{subfigure}
\usepackage{amsthm}
\usepackage{threeparttable}

\newcommand\ergs{\,erg\,s$^{-1}$}

\newcommand{\kev}{\,keV}

\newcommand{\psqcm}{\,cm$^{-2}$}
\newcommand{\ps}{\,s$^{-1}$}

\newcommand{\erg}{\,erg}

\newcommand{\err}[3]{$#1_{-#2}^{+#3}$}

\makeatletter
\providecommand\@onlyoneemail{}
\providecommand\@authoremail{}

\newcommand{\correspondauthor}[2]{
  \ifx\@authoremail\@empty
    \def\@authoremail{#1, email: #2}
    \def\@onlyoneemail{}
  \else
    \def\@onlyoneemail{No}
    \expandafter\def\expandafter\@authoremail\expandafter{\@authoremail; #1, email: #2}
  \fi
}

\makeatother
\makeatletter
\renewcommand{\Authorfootnote}{%
  \par
  \begin{flushleft}
    \renewcommand{\baselinestretch}{0.7}%
    \noindent\rule{2.5cm}{0.4pt}\\[0.1mm]
    \qihao
    \ifx\@onlyoneemail\@empty
      *Corresponding author (email: \@authoremail)%
    \else
      *Corresponding authors (\@authoremail)%
    \fi
    \ifx\@contributions\@empty\else
      \\\noindent\dag{\thinspace}\@contributions
    \fi
  \end{flushleft}
}
\makeatother

\graphicspath{{./}{fig/}}

\begin{document}

\ensubject{subject}

\ArticleType{Article}
\SpecialTopic{SPECIAL TOPIC: }
\Year{2023}
\Month{January}
\Vol{66}
\No{1}
\DOI{??}
\ArtNo{000000}
\ReceiveDate{January 11, 2023}
\AcceptDate{April 6, 2023}

\title{Anisotropic Particle Transport from a Pulsar Wind Nebula Revealed by Einstein Probe and LHAASO}{Anisotropic Particle Transport from a Pulsar Wind Nebula Revealed by Einstein Probe and LHAASO}

\author[1,2,3]{Zhen~Cao}
\author[3,4,5,6]{F.~Aharonian}
\author[1,3]{Y.X.~Bai}
\author[7]{Y.W.~Bao}
\author[8]{D.~Bastieri}
\author[1,2,3]{X.J.~Bi}
\author[1,3]{Y.J.~Bi}
\author[7]{W.~Bian}
\author[9]{J.~Blunier}
\author[10]{A.V.~Bukevich}
% ...

\AuthorMark{Cao Z}
\AuthorCitation{Cao Z et al}

\affil[1]{State Key Laboratory of Particle Astrophysics \& Experimental Physics Division \& Computing Center, Institute of High Energy Physics, Chinese Academy of Sciences, 100049 Beijing, China}
\affil[2]{University of Chinese Academy of Sciences, 100049 Beijing, China}
\affil[3]{Tianfu Cosmic Ray Research Center, 610000 Chengdu, China}
\affil[4]{University of Science and Technology of China, 230026 Hefei, China}
\affil[5]{Yerevan State University, 1 Alek Manukyan Street, Yerevan 0025, Armenia }
\affil[6]{Max-Planck-Institut for Nuclear Physics, P.O. Box 103980, 69029  Heidelberg, Germany}
\affil[7]{Tsung-Dao Lee Institute \& School of Physics and Astronomy, Shanghai Jiao Tong University, 200240 Shanghai, China}
\affil[8]{Center for Astrophysics, Guangzhou University, 510006 Guangzhou, China}
\affil[9]{APC, Universit'e Paris Cit'e, CNRS/IN2P3, CEA/IRFU, Observatoire de Paris, 119 75205 Paris, France}
\affil[10]{Institute for Nuclear Research of Russian Academy of Sciences, 117312 Moscow, Russia}
% ...

\abstract{Pulsar wind nebulae (PWNe) are major cosmic ray accelerators,  yet the mechanisms transporting high-energy particles into the interstellar medium remain elusive. Building on the LHAASO discovery of an ultra-high-energy (UHE) $\gamma$-ray source near the bow-shock PWN powered by the pulsar PSR~J1740+1000, we present a joint Einstein Probe (EP) and LHAASO study of this system. EP observations reveal an extended X-ray tail far exceeding the structure previously seen by XMM-Newton. Updated LHAASO observations show that the $\gamma$-ray emission is elongated, with its major axis aligned with the extended X-ray tail revealed by EP. This is the first detection of an X-ray pulsar tail associated with a spatially coincident extended UHE $\gamma$-ray emission. The X-ray and $\gamma$-ray spectrum can be well explained with a single population of relativistic electrons via synchrotron and inverse Compton radiation, respectively,  removing the need for particle re-acceleration during propagation. The results unambiguously show that electrons/positrons above 100\,TeV are escaping from the PWN. Instead of the immediate, isotropic diffusion into ambient interstellar medium that is typically assumed, these particles are transported anisotropically over at least $\sim$10\,pc, either guided by the background magnetic field or carried by an advective outflow.
}



\maketitle

\correspondauthor{Y.-H. Chi}{yhchi@smail.nju.edu.cn}
\correspondauthor{H. Feng}{hfeng@ihep.ac.cn}
\correspondauthor{R.-Y. Liu}{ryliu@nju.edu.cn}
\correspondauthor{S.-Q. Xi}{xisq@ihep.ac.cn}
\correspondauthor{P. Zhou}{pingzhou@nju.edu.cn}

\begin{multicols}{2}
\def\floatpagepagefraction{1}
\def\textpagefraction{.001}

\section{Introduction} \label{sec:intro}

Pulsar wind nebulae (PWNe) represent one of the most efficient classes of Galactic particle accelerators, converting the rotational energy of the central pulsar into broad-band non-thermal emission from radio to ultrahigh energy (UHE; $>100$\,TeV) $\gamma$-rays \cite{Cao2021_12UHE}. The radio-to-X-ray spectrum is dominated by synchrotron radiation from relativistic electrons accelerated at the pulsar wind termination shock or upstream of the pulsar wind, while $\gamma$-ray emission (GeV to PeV) may arise from inverse Compton (IC) scattering by these electrons off ambient photon fields (e.g., the cosmic microwave background (CMB), interstellar radiation field (IRB), and/or its own synchrotron radiation), or possibly from hadronic processes involving relativistic protons interacting with ambient matters \cite{2006ARA&A..44...17G}.

PWNe are ideal places to study the transport of cosmic rays in the parsec scale. Since pulsars may acquire high natal kick velocities, they may escape their natal supernova remnants and interact directly with the interstellar medium (ISM) as they age. When the pulsar's proper motion exceeds the local sound speed, it drives a bow shock, confining the pulsar wind into a cometary tail structure known as a bow-shock pulsar wind nebula (BSPWN). These systems are typically observed in X-rays as collimated synchrotron-emitting tails trailing the pulsar and extending to several parsecs \cite{Kargaltsev17, Bykov2017}. On the other hand, X-ray filamentary or jet-like structures are observed strongly misaligned with the PWN tail in a few systems like the Guitar Nebula \cite{Hui2007} and Lighthouse Nebula \cite{Pavan2014, Pavan2016}. \Authorfootnote Extending far beyond the pulsar's immediate vicinity, these elongated features are interpreted as signatures of high-energy particles escaping along ordered magnetic field lines or via collimated outflows \cite{Bandiera08}. 
The recent discovery of extended TeV halos around middle-aged pulsars has highlighted the efficient escape and propagation of high-energy electrons/positrons from PWNe \cite{Abeysekara2017, Aharonian2021}. 
However, the mechanism by which particles accelerated within the PWN are transported through the tail or filamentary structures into the ambient ISM remains poorly understood.

\begin{figure*}[ht!]
    \centering
    \includegraphics[width=\textwidth]{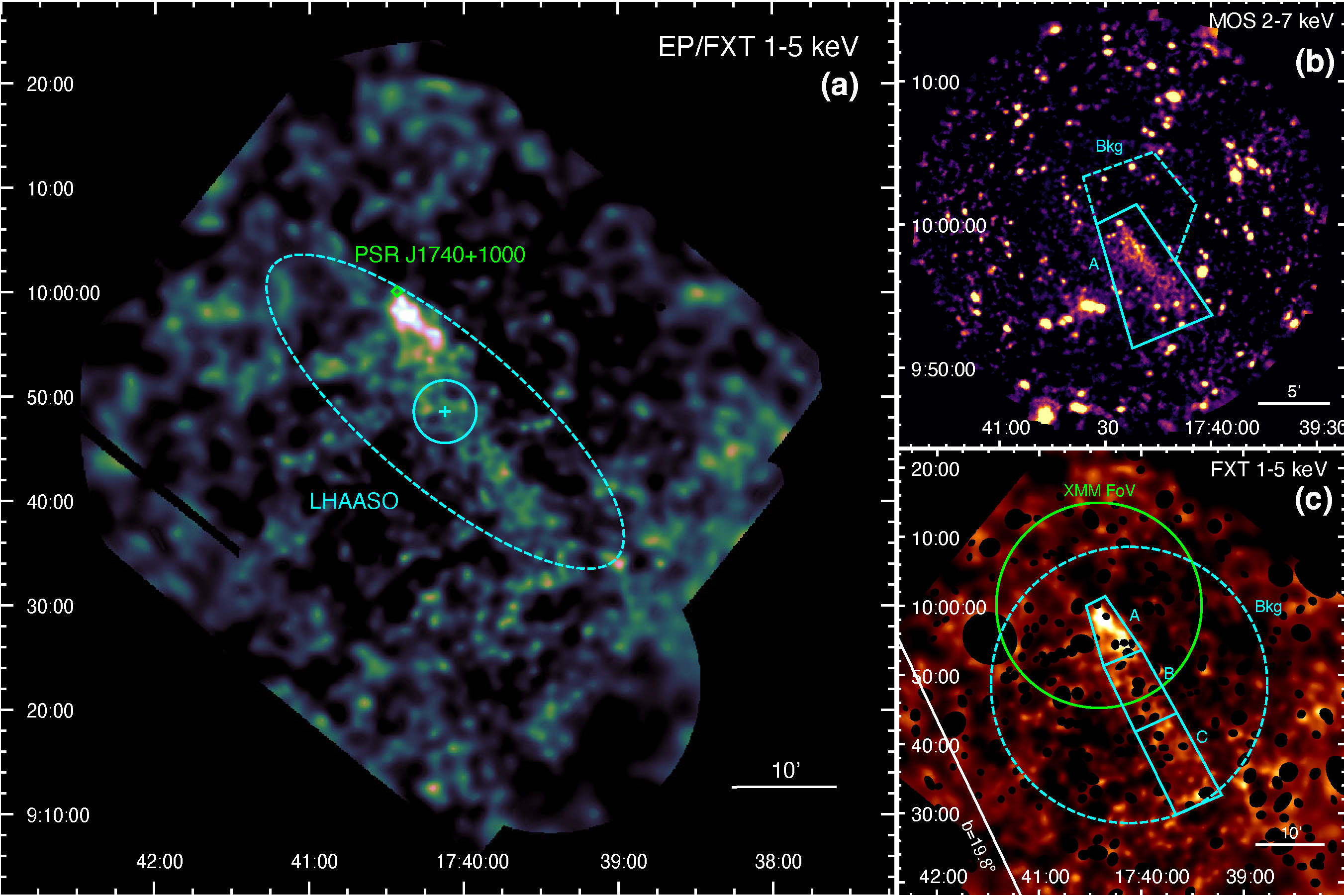}
    \caption{X-ray images around the pulsar PSR~1740+1000 in celestial coordinators. \textbf{(a)}: Point-like sources removed, background subtracted, vignetting corrected, adaptively smoothed FXT image in the 1--5 keV band, overlapped with the LHAASO $\gamma$-ray position (cross), position uncertainty (solid line), and $2\sigma$ elliptical extension (dashed line). The position of PSR~1740+1000 is marked in green. \textbf{(b)}: XMM-Newton 2--7\,keV image with point-like sources remained. The solid lines encircle the source extraction region and the dashed lines encircle the background region. \textbf{(c)}: same as \textbf{(a)} but with point-like sources masked. The solid lines encircle the three source regions marked as A, B, and C, respectively, and the dashed circle (excluding the source regions) encircles the sky background region. The green solid circle marks the field of view of XMM MOS shown in \textbf{(b)}. The white solid line indicates the Galactic latitude at $b = 19.8^\circ$. The horizontal lines indicates the scale bars labeled with the angular sizes. }
    \label{fig: mor}
\end{figure*}

The middle-aged pulsar PSR~J1740+1000 ($\tau \sim 114$\,kyr, $\dot{E} = 2.3\times 10^{35}$\ergs), located at a high Galactic latitude ($b \sim 20^\circ$) and a distance of $\sim$1.4\,kpc, powers a well-studied BSPWN. Previous XMM-Newton observations revealed an X-ray synchrotron tail extending $\sim$$6^\prime$ to the southwest of the pulsar \cite{Kargaltsev08}. Recently, the Large High Altitude Air Shower Observatory (LHAASO) discovered a point-like UHE $\gamma$-ray source, LHAASO~J1740+0948, with a significance of $17.1\sigma$ above 25\,TeV near the PWN \cite{Cao2025_1740}. Intriguingly, the $\gamma$-ray emission is significantly offset from both the pulsar and the previously identified X-ray tail, but seems to align with the tail's extrapolated trajectory further downstream. This may suggest that there is an efficient mechanism for transporting high energy particles from the PWN to the ISM, or the occurrence of particle re-acceleration during propagation, contradicting conventional pulsar halo models. 

The spatial misalignment between the X-ray tail and the $\gamma$-ray emission poses significant theoretical challenges. 
An earlier study \cite{Benbow21} predicted the TeV $\gamma$-ray flux of PWN~J1740+1000 based on its X-ray emission. This predicted flux is lower than the level observed by LHAASO \cite{Cao2025_1740}. Although the model may be reconciled with observation given a weaker magnetic field within the tail than expected, it is not clear whether the $\gamma$-ray source detected by LHAASO has the same origin with the X-ray BSPWN detected by XMM-Newton. Recently, several attempts \cite{Brunelli25, Gagnon25} have been made to search for X-ray emission associated with the UHE source. While no significant diffuse emission has been confirmed, Brunelli et al. \cite{Brunelli25} reported a marginally enhanced hard X-ray emission near the position of the LHAASO source; Gagnon et al. \cite{Gagnon25} analyzed nearby point-like sources, arguing against them being the origin of the cosmic rays. The Follow-up X-ray Telescope (FXT \cite{Chen20}) onboard the Einstein Probe (EP) satellite \cite{Yuan22} offers a unique opportunity to search for such diffuse, faint synchrotron X-rays, thanks to its large $\sim1^\circ$ field of view and exceptionally low instrumental particle background.

In this paper, we present a combined analysis of the PSR J1740+1000 system using deep X-ray observations with EP-FXT and $\gamma$-ray observations with LHAASO. Our EP-FXT observations reveal that the pulsar tail extends significantly further than previously known, reaching the exact location of the UHE emission.  
Combined with updated LHAASO data, the morphologies and spectral energy distribution (SED) across both bands can be consistently explained by a single population of high energy electrons, providing key insights into the injection of high energy cosmic rays into the ISM.

\section{X-ray imaging and spectroscopy}
\label{sec:x-ray}

EP observed the region around the $\gamma$-ray source with a total exposure of about 71\,ks. 
To enable the detection of faint, diffuse emission, we removed point-like sources, and applied the vignetting correction and adaptive smoothing (see details in the online
supplementary material).  
The resultant  EP-FXT image in the 1--5 keV energy band around the LHAASO source is displayed in Figure~\ref{fig: mor}. 
As one can see, an X-ray filament is visible extending towards southwest of the pulsar and spatially coincident with the LHAASO $\gamma$-ray emission peak. 
Such a structure is reminiscent of a BSPWN and we name it ``extended tail'' as it extends up to $\sim$$32'$ away from the pulsar, corresponding to $\sim$13\,pc at 1.4\,kpc. In regions farther out, the diffuse emission cannot be significantly distinguished from the background. This extended tail has a nearly uniform width of $\sim$$6'$ and is roughly aligned with the Galactic longitude direction within a few degrees. 

With a deep exposure of $\sim$450\,ks, XMM-Newton revealed fine structures in part of the PWN (Figure~\ref{fig: mor}c), the brightest, northern portion of the extended tail. We selected the 2--7\,keV band of the XMM-Newton data to highlight the non-thermal synchrotron emission, while at lower energies the photons are contaminated by prominent instrument lines and dominated by thermal emission. The majority of the flux comes from a straight filamentary tail, extending $\sim$$8'$ (3.3\,pc) from PSR J1740+1000 to the southwest. Another shorter, filamentary branch appears to extend southward, with a length of $\sim$$2.6'$ (1.0\,pc). Rather than originating at the pulsar position, it connects with the brighter filament at its northern end.

We divided diffuse X-ray emission into three regions A, B, and C (Figure~\ref{fig: mor}c) to study the variation of spectral properties along the extended tail. Region A corresponds to the X-ray tail observed by XMM-Newton with a length of $\sim8'$, allowing for a direct comparison between the XMM-Newton and FXT. The more extended part is divided into region B and C with equivalent length.

\begin{table*}[ht]
    \centering
    \caption{X-ray spectral fitting results of the diffuse emission in the three regions.}
    \renewcommand{\arraystretch}{1.5}
    \begin{tabular*}{\textwidth}{@{\extracolsep{\fill}}cllll}
    \noalign{\smallskip}\hline
        Region & A (MOS2) & A (FXT) & B & C \\ \hline
        $\Gamma$ & \err{1.62}{0.04}{0.04} & \err{1.43}{0.16}{0.16} & \err{1.59}{0.20}{0.21} & \err{1.78}{0.28}{0.30} \\
        $\log F^a$ & \err{-12.49}{0.01}{0.01} & \err{-12.47}{0.04}{0.04} & \err{-12.53}{0.05}{0.05} & \err{-12.49}{0.06}{0.06} \\
        $\log F_{\rm l.l.}^b$ & $-12.54$ & $-12.68$ & $-12.74$ & $-12.74$ \\
        $\chi^2/{\rm dof}$ & 674.3/595 & 32.6/35 & 28.1/27 & 20.0/24 \\ \hline
        Flux Density$^c$ & $2.73\pm0.06$ & $2.85\pm0.26$ & $1.46\pm0.17$ & $1.13\pm0.16$ \\
        $F_{\rm diff}^d$ & $4.05\pm0.09$ & $4.24\pm0.39$ & $3.51\pm0.40$ & $3.77\pm0.52$ \\ \hline\noalign{\smallskip}
    \end{tabular*}
    \begin{tablenotes}
    \item Note: the uncertainties are quoted at 1$\sigma$ confidence.
        \item $^a$Unabsorbed 0.5--8\kev\ flux in units of erg\ps\psqcm.
        \item $^b$$5\sigma$ lower limit of the flux in 0.5--8\kev.
        \item $^c$Unabsorbed flux density in units of $10^{-18}$\erg\ps\psqcm\,arcsec$^{-2}$. \item $^d$Corrected flux of diffuse emission in units of $10^{-13}$\erg\ps\psqcm.
    \end{tablenotes}
    \label{tab: result1740-new}
\end{table*}

\begin{figure*}
\centering
\includegraphics[width=0.95\textwidth]{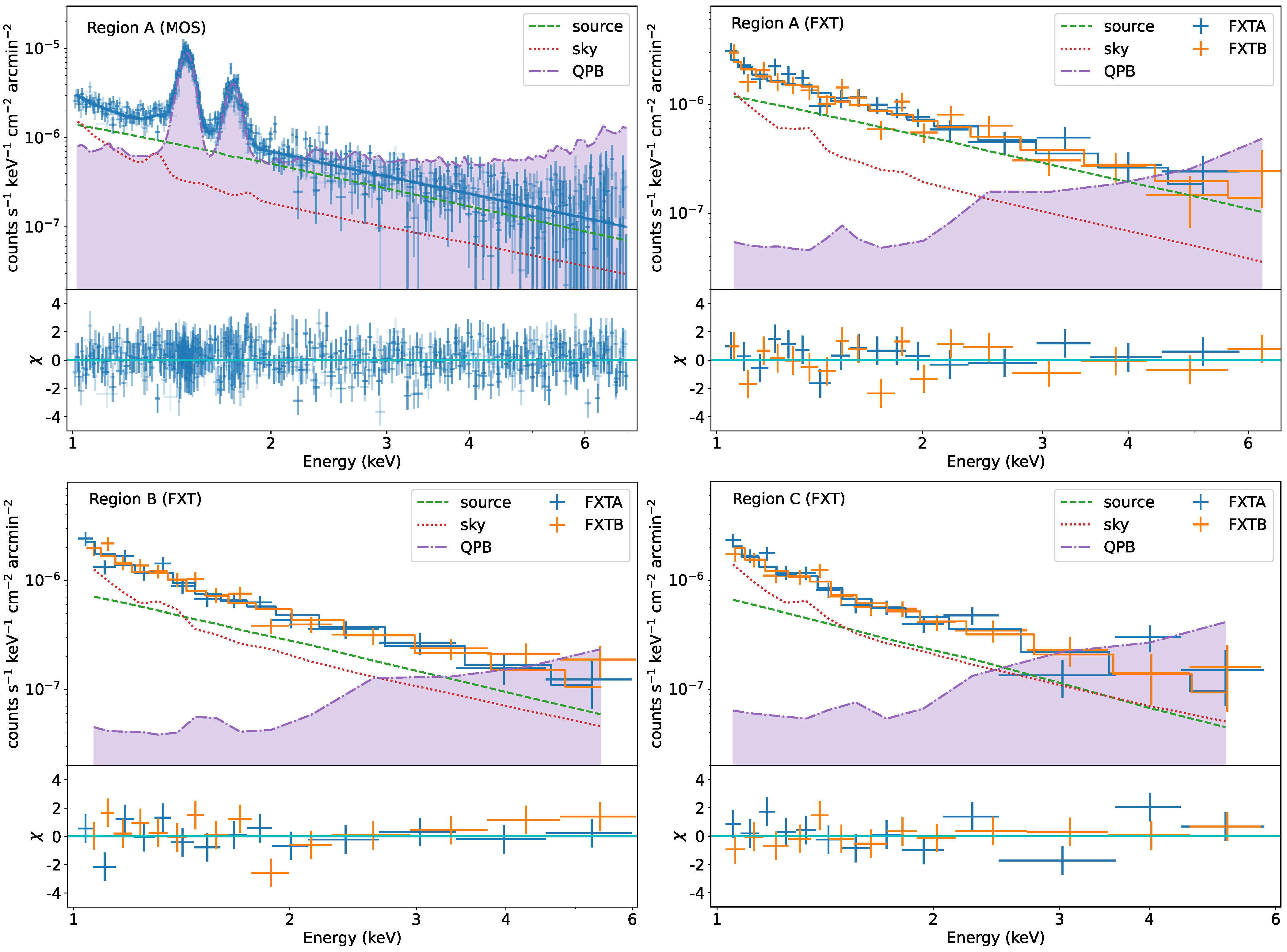}
\vspace{-10pt}
\caption{XMM-Newton and EP spectra in different regions with best-fit models. The green dashed line indicates the synchrotron emission from the source, the red dotted line indicates the sky background, and the purple dash-dotted line indicates the instrumental quiescent particle background (QPB).}
\vspace{-35pt}
\label{fig:xrayspec}
\end{figure*}

The fitting results are listed in Table~\ref{tab: result1740-new}, and the corresponding spectra are plotted in Figure~\ref{fig:xrayspec}. 
EP-FXT has a significantly lower particle background than XMM-Newton. In region A, the particle background of XMM-Newton exceeds the source emission at around 2--3\,keV, whereas for FXT, this crossover occurs above 4 keV. The best-fit parameters (i.e., the photon index and flux) derived between FXT and XMM-Newton are consistent with each other within 1$\sigma$ uncertainties.

The photon index is generally consistent with the values ($\sim$1.4--1.8) reported in previous X-ray studies \cite{Kargaltsev08, Benbow21, Brunelli25, Gagnon25}. However, the measured flux in region A (approximately $4\times10^{-13}$\,erg\,s$^{-1}$\,cm$^{-2}$) appears to be a factor of 2--3 higher. This discrepancy arises because our analysis covers a larger region than that used in previous studies, who limited their analysis in the central MOS region that is smaller than region A. 

The low instrumental background of FXT allows us to unveil a significant contribution of fluxes from regions B and C. Compared to region A, their surface brightness is $\lesssim50\%$ lower, yet their integrated fluxes are comparable. The total unabsorbed flux from all three regions is $1.1\times10^{-12}$\,erg\,s$^{-1}$\,cm$^{-2}$. At a distance of 1.4\,kpc, this corresponds to a 0.5--8\,keV luminosity of $\sim2.6\times10^{32}$\,erg\,s$^{-1}$, implying a radiative efficiency (i.e., the ratio of X-ray luminosity to spin-down luminosity) of $\gtrsim10^{-3}$, which is much higher than that reported in previous studies \cite{Kargaltsev08}. Among the three regions, the photon index shows no significant spatial variation, all consistent within 1$\sigma$ uncertainties.

\section{Gamma-ray morphology and spectroscopy}
\label{sec:lhaaso}

\begin{figure*}
    \centering
    \includegraphics[width=0.95\textwidth]{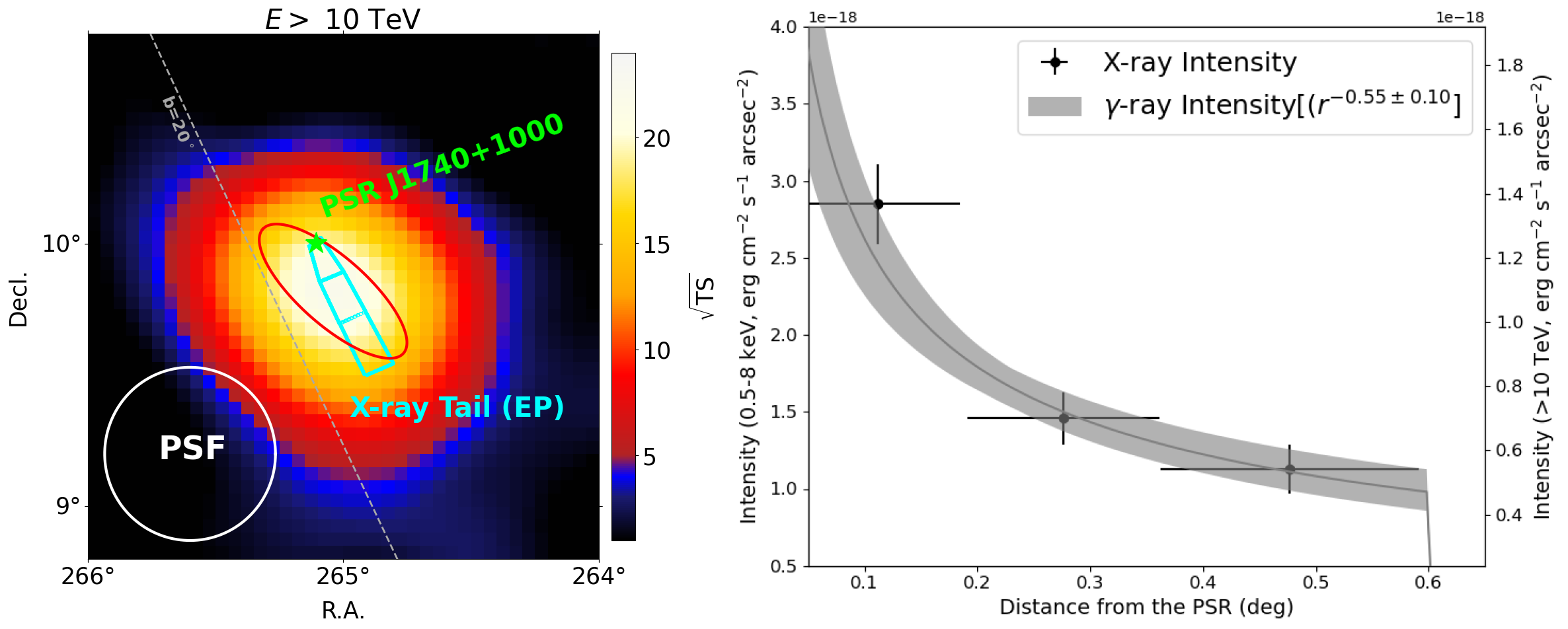}
    \vspace{-10pt}
    \caption{Left: significance map of LHAASO J1740+0948 at energies above 10 TeV. Overlaid are the extended X-ray tail in cyan. Lime star represents the position of the pulsar PSR J1740+1000. The red ellipse represents a semi-major axis of 0.36$^\circ$ and a semi-minor axis of 0.12$^\circ$. The orientation angle is 49.6$^\circ$ east of north. Right: surface brightness as a function of the angular distance from the pulsar position. Points are measured X-ray data and the gray curve and region indicate the best-fit $\gamma$-ray profile.}
    \label{fig:lhaasosig}
\end{figure*}

The significance map of LHAASO J1740+0948 at energies above 10 TeV is shown in ‌Figure~\ref{fig:lhaasosig}. The pulsar PSR J1740+1000 is clearly offset from the weighted centroid of the $\gamma$-ray emission, as reported in Cao et al.\ \cite{Cao2025_1740}, which locate at the extended X-ray tail detected with FXT.

Updated LHAASO data suggest that the $\gamma$-ray morphology can be better described by an elliptical Gaussian than a point source at high confidence, with an elongation consistent with the orientation of the extended tail. We adopt the 2$\sigma$ extension along the semi-major axis of $\sim0.36^\circ$ and the $95\%$ confidence level upper limit for the semi-minor axis of $\sim0.12^\circ$ to describe the size of the $\gamma$-ray emission in this study.
We also constructed a template based on the X-ray emission of the extended tail to fit the $\gamma$-ray morphology, and obtained a slightly better but comparable fit (see the online supplementary materials).

In the leptonic framework, the $\gamma$-ray morphology closely traces the distribution of electrons, given a homogeneous target photon field over the source region such as CMB and IRB. To investigate the surface density distribution of relativistic electrons along the X-ray tail, we assume the surface brightness of $\gamma$-ray emission to follow a power-law distribution defined as $S \propto r^{-\alpha}$, where $r$ is the distance to the pulsar. The $\gamma$-ray emission region is assumed to be the same as the X-ray extension region, i.e., a width of $6'$ and a length of $32'$. To avoid the singularity of the power-law distribution, the fitting region begins at $0.05^\circ$ from the pulsar. We can effectively constrain the parameter to $\alpha = 0.55 \pm 0.10$. The scenario where $\alpha=0$ corresponds to a uniform electron distribution along the tail, which is ruled out at a 4.3$\sigma$ ($\Delta {\rm TS} = 18.4$) confidence level. As shown in the right panel of Figure~\ref{fig:lhaasosig}, we observe a good correlation between the surface brightness distribution of X-ray and that of $\gamma$-ray along the tail. To quantify the correlation between the X-ray and $\gamma$-ray morphologies, we assume the $\gamma$-ray intensity profile scales with the X-ray intensity profile by $S_\gamma\propto r^{-\beta} S_x$,
and obtain $\beta = 0.12 \pm 0.13$.

The photon spectrum is presented in Figure~\ref{fig:sed_lhaaso_ep}, which can best-fitted with a log-parabola model with an integral energy flux $F({>10\,\mathrm{TeV}}) \approx  5.5 \times 10^{-13}\,\mathrm{erg\,cm^{-2}\,s^{-1}}$. 

\section{Discussions and Conclusions} \label{sec:discussion}

Updated LHAASO data suggest that LHAASO J1740+0948 is not a point-like source, but exhibits an extension aligned with the extended X-ray tail. The spatial consistency between the X-ray and $\gamma$-ray emission, combined with the decreasing $\gamma$-ray surface brightness $S_{\gamma}\propto
r^{-0.55}$ away from the pulsar, relaxes the requirement of particle re-acceleration in the tail of BSPWN as discussed in the previous study \cite{Cao2025_1740}, although the possibility cannot be excluded yet. 
Instead, in accordance with Occam's Razor, we favor the simplest explanation that the emission in both bands originates from the same population of electrons, via synchrotron radiation in X-rays and IC radiation in $\gamma$-rays. 
Since the target photon field for IC radiation is basically known and homogeneous over the source region, the spectrum of parent electrons can be effectively constrained by the $\gamma$-ray observation,  allowing for a further determination of the magnetic field strength through a multiwavelength SED modeling.

\begin{figure}[H]
    \centering
    \includegraphics[width=\columnwidth]{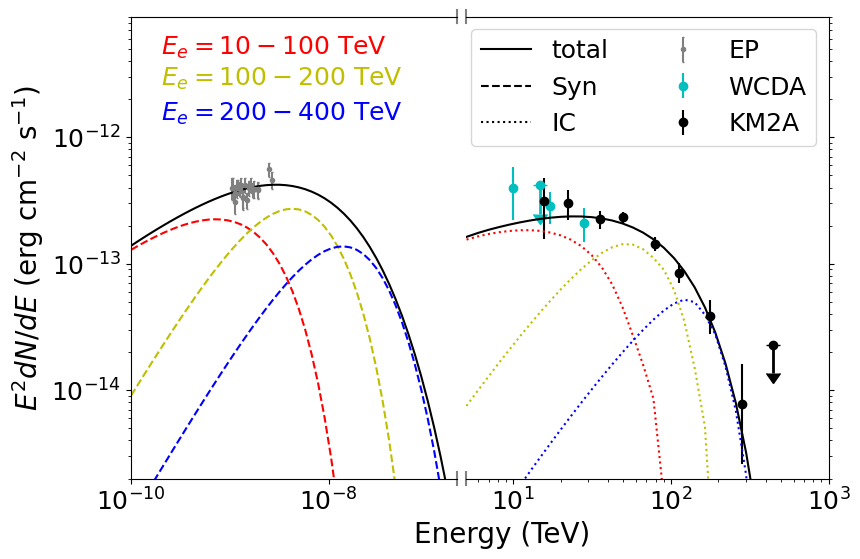}
    \caption{Multiband SED of LHAASO J1740+0948 and spectral modeling. The dashed and dotted lines represent synchrotron and IC radiation, respectively, by the same population of electrons at a certain energy range (denoted by different colors). }
    \label{fig:sed_lhaaso_ep}
\end{figure}

We assume the steady-state electron spectrum following a power-law distribution terminated by a super-exponential cutoff:
\begin{equation}\label{splc}
    \frac{dN_{\rm e}}{dE_{\rm e}} \propto E_{\rm e}^{-\alpha} \exp\left[-\left(\frac{E_{\rm e}}{E_{\rm c}}\right)^2\right] \, ,
\end{equation}
where $E_{\rm e}$ is the primary particle energy, $\alpha$ is the spectral index, and $E_{\rm c}$ is the exponential cutoff energy. The photon field contains the CMB as well as an IRB described with the far-infrared radiation ($T_{\rm FIR} = 33$\,K and energy density $u_{\rm FIR} = 0.27 \,\rm eV/cm^3$) and the starlight ($T_{\rm star} = 4700$\,K and energy density $u_{\rm star} = 0.49 \,\rm eV/cm^3$) based on the interstellar radiation field model \cite{Popescu2017}.
Using the open-source package GAMERA\footnote{\url{https://libgamera.github.io/GAMERA/docs/tutorials\_main.html}}, an adequate fit is obtained with $\chi^2 = 17.4$ and ${\rm dof} = 30$, and the SED modeling is presented in Figure~\ref{fig:sed_lhaaso_ep}. 
The magnetic field strength is constrained to $3.0\pm0.2\ \mu$G, which is consistent with the typical interstellar magnetic field.

Note that since the target radiation field is homogeneous, we expect $S_\gamma/S_{\rm X}\propto B(r)^{-2}$, leading to $B(r)\propto r^{0.06\pm 0.07}$. 
The magnetic field exhibits negligible spatial variation. It is possible that the extended tail is produced by freshly escaping particles which diffuse along the interstellar magnetic field lines \cite{Liu2019, Bao2025a, Bao2025b}, which is similar to the peanut-shaped $\gamma$-ray source consisting of LHAASO J0216+4239 and LHAASO J0207+4300 recently reported by LHAASO \cite{Cao2025_peanut} and the so-called ``misaligned outflow'' seen in some other BSPWN (see \cite{Kargaltsev17} and references therein). 
In this scenario, the diffusion coefficient need be sufficiently large so that electrons can diffuse over the observed (projected) length (i.e., 13\,pc) before being cooled, provided that no re-acceleration of particles taking place in the tail. For 100\,TeV electrons, which are responsible for keV synchrotron radiation and $\sim 10$\,TeV IC radiation, the cooling timescale is about 10\,kyr in the $3\,\rm \mu G$ magnetic field and the considered radiation field. 
Therefore, the diffusion coefficient along the magnetic field can be expressed as

\begin{equation}\label{eq:D_condition}
\begin{split}
    D_{\|} \geq & \frac{(l/\sin\theta)^2}{2t_{\rm cool}} \\
    \approx & 10^{28}\,\mathrm{cm^2\,s^{-1}}\,\left(\frac{l}{13\,\mathrm{pc}}\right)^2\left(\frac{\sin\theta}{0.5} \right)^{-2}\left(\frac{t_{\rm cool}}{10\,\rm kyr}\right)^{-1} \, ,
\end{split}
\end{equation}
where $\theta$ is the angle between the observer's line of sight and the true direction of the tail. This lower limit of diffusion coefficient is about two orders of magnitude lower than the standard diffusion coefficient in the ISM for 100\,TeV particles, similar to that inferred within typical pulsar halos \cite{Abeysekara2017, Aharonian2021}. The length of the tail may reflect the diffusion length constrained by cooling of electrons if the condition of Eq.~(\ref{eq:D_condition}) is marginally satisfied. Alternatively, the length of the tail may reflect the coherence length of the magnetic field, beyond which the mean magnetic field changes its direction so that escaping electrons are guided to other directions or become more diffusive.

Alternatively, one cannot rule out the possibility that the tail is produced by confined electrons advected with the downstream flow of the BSPWN. In this scenario, this tail is the longest X-ray PWN tail known to date, significantly exceeding the 7\,pc tail of PSR J1509$-$5850 at 3.8\,kpc \cite{Klinger16, Kargaltsev17}. The average flow velocity needs be
\begin{equation}
\begin{split}
    v_{\rm flow} \geq & \frac{l/\sin\theta}{t_{\rm cool}} \\
    \approx & 2700\,\mathrm{km\,s^{-1}} \left(\frac{l}{13\,\mathrm{pc}}\right)\left(\frac{\sin\theta}{0.5}\right)^{-1}\left(\frac{t_{\rm cool}}{10\,\rm kyr}\right)^{-1} \, ,
\end{split}
\end{equation}
which is similar to the estimated outflow velocity seen from BSPWNs like PSR B0355$+$54, PSR J1509$-$5850, and PSR J1101$-$6101 \cite{Kargaltsev17}. This velocity is much higher than the proper motion velocity of PSR~J1740+1000 ($\lesssim60\,\mathrm{mas\,yr^{-1}}$ or $\lesssim400\,\mathrm{km\,s^{-1}}$ \cite{Halpen13}), indicating that the tail structure is not produced by the proper motion of the pulsar but due to much faster outflows ejected from the accelerator. For both scenario, the tail-like morphology indicates that the transport of the relativistic electrons perpendicular to the magnetic field is suppressed, which is likely due to a highly ordered magnetic field.

If the intensity profiles and photon index profiles in both X-ray and $\gamma$-ray bands are precisely measured, we may distinguish the two scenarios by comparing the theoretical predictions and measurements. In either scenario, nevertheless, our study suggests that when high-energy particles leave their accelerators, they do not immediately diffuse into the surrounding medium. 
Instead, these particles are likely transported a significant distance from the acceleration site --- either via anisotropic propagation along the background magnetic field or through advective outflow --- prior to undergoing the isotropic diffusion typically assumed in the literature.
Such a behavior may elucidate the spatial offsets observed between some $\gamma$-ray sources and their candidate astrophysical counterparts \cite{Cao24}, and may also lead to a different $\gamma$-ray intensity profile close to the accelerator from the prediction under the pure diffusion scenario. 

To conclude, in this work we present the discovery of a 13\,pc extended tail, a BSPWN powered by PSR~J1740+1000, with EP and LHAASO observations. 
The morphological alignment and spectral consistency between the X-ray and $\gamma$-ray emission suggest that both originate from the same population of relativistic electrons, via synchrotron and inverse Compton processes, respectively, without the need for particle re-acceleration.
The inferred magnetic field strength is about 3\,$\mu$G; no spatial variation along the tail is found. 
This result, together with the observed surface brightness profiles, supports a scenario in which high-energy electrons/positrons escape the PWN and undergo parsec-scale transport, either via anisotropic diffusion along ordered magnetic field lines or through an advective outflow confined within the tail structure.
Both mechanisms imply that particles do not immediately isotropize upon leaving the accelerator; instead, they undergo a phase of coherent propagation over a significant distance before mixing into the ``sea'' of cosmic rays.
These findings are not possible without the combination of the low background and large field of view of EP-FXT and the unprecedented sensitivity of LHAASO. 

\Acknowledgements{We appreciate the valuable discussion on the X-ray sky background with Gabriele Ponti and the comments from the anonymous referee. This research work is supported by the following grants: The National Natural Science Foundation of China Nos. 12393852, 12273010, 12025301, 12393851, 12393853, 12393854, 12205314, 12105301, 12305120, 12261160362, 12105294, U1931201, 12375107, 12173039, the Department of Science and Technology of Sichuan Province, China No. 24NSFSC2319, Project for Young Scientists in Basic Research of Chinese Academy of Sciences No.YSBR-061, and in Thailand from the NSRF via the Research and Innovation Acceleration Agency for Competitiveness and Area Development (RCAD) (Program Management Unit for Technology and Innovation for Future Industries (PMU-B): Brainpower for Future Industries) [grant number B39G690003]. Y.H.C. and P.Z. acknowledge the support from the Fundamental Research Funds for the Central Universities with grant No. KG202502. This work is based on the data obtained with Einstein Probe, a space mission supported by the Strategic Priority Program on Space Science of Chinese Academy of Sciences, in collaboration with the European Space Agency, the Max-Planck-Institute for extraterrestrial Physics (Germany), and the Centre National d'Études Spatiales (France). We would like to thank all staff members who work at the LHAASO site above 4400 meter above the sea level year round to maintain the detector and keep the water recycling system, electricity power supply and other components of the experiment operating smoothly. We are grateful to Chengdu Management Committee of Tianfu New Area for the constant financial support for research with LHAASO data. We appreciate the computing and data service support provided by the National High Energy Physics Data Center for the data analysis in this paper. This study is based on observations obtained with XMM-Newton, an ESA science mission with instruments and contributions directly funded by ESA Member States and NASA.}

\InterestConflict{The authors declare that they have no conflict of interest.}

\Supplements{Supplementary materials to this article can be found online.}

\bibliographystyle{scpma}
\bibliography{J1740}

\end{multicols}
\end{document}


\maketitle

\section{EP/FXT}
\label{sec:ep}

\subsection{Observations and data reduction}
FXT observed the center of 1LHAASO J1740+0948u with two epochs, and the southwestern region with four short epochs, with a total exposure of $\sim70$\,s and a sky coverage of 1.7 degree$^2$.
The observations were conducted in the full frame mode with the Thin filter. The detailed information is listed in Supplementary Table~\ref{tab: obs}. The process and analysis of the FXT data were mainly based on the FXT Data Analysis Software (FXTDAS, ver.\ 1.20 with FXT CALDB ver.\ 1.20), and HEASoft (ver.\ 6.35 \cite{heasoft14}) and CIAO (ver.\ 4.16 \cite{Fruscione06}) were also utilized.

\begin{table*}
    \centering
    \caption{EP/FXT observations. }
    \begin{tabular}{ccccccc}
    \hline
        Pointing & Obs. ID & Obs. Date & \multicolumn{2}{c}{Exposure (ks)} & \multicolumn{2}{c}{Rate$^a$ (cts\,s$^{-1}$)} \\
        ~ & ~ & ~ & FXTA & FXTB & FXTA & FXTB \\ \hline
		1LHAASO J1740+0948u & 11904194387 & 2024-03-24 & 11.9 & 11.9 & 3.0 & 4.0 \\
		1LHAASO J1740+0948u & 11904194561 & 2024-07-11 & 41.5 & 41.3 & 2.6 & 3.3 \\
        LHAASO J1740-SW & 11900263680 & 2025-06-08 & 4.7 & 4.2 & 2.4 & 3.3 \\
        LHAASO J1740-SW & 11900269952 & 2025-06-14 & 2.2 & 2.0 & 2.4 & 3.7 \\
        LHAASO J1740-SW & 11900279936 & 2025-06-22 & 7.7 & 7.4 & 2.3 & 3.2 \\
        LHAASO J1740-SW & 11900286464 & 2025-06-27 & 2.6 & 2.5 & 2.3 & 3.2 \\ \hline
        Total & ~ & ~ & 70.6 & 69.3 \\
        \hline
    \end{tabular}
    \begin{tablenotes}
        \item $^a$counts rate of the entire field of view in 0.4--7\,keV.
    \end{tablenotes}
    \label{tab: obs}
\end{table*}

The FXT data are reprocessed through a standard sequence of steps, including coordinate update ({\it fxtcoord}), gain calibration ({\it fxtpical}), removal of particle-induced background ({\it fxtparticleidentify}), identification of bad, hot, and flickering pixels ({\it fxtbadpix} and {\it fxthotpix}), events reconstruction ({\it fxtgrade}), and good time interval (GTI) selection ({\it fxtgtigen}). This entire workflow can also be executed via a single command {\it fxtchain}. The events are then filtered with {\it xselect}. To remove time epochs contaminated by bright Earth and background flares (e.g., \cite{Kuntz08}), we extract the light curve in 0.4--7\kev\ band of the entire field of view with $\sim100$ counts in each bin and preclude the bins with count rates 40\% higher than the median.
Meanwhile, we compare the average counts rate of the filtered events file among all the observations (listed in Supplementary Table~\ref{tab: obs}). The observations 11904194387 and 11900269952 have a higher FXTB counts rate, indicative of a long-term contamination of soft protons. Therefore we do not use them for further analysis.

\subsection{Imaging analysis}

We combine all the events by projecting them onto a common tangent point and extracted the image with a pixel size of 4$^{\prime\prime}$.
The FXT instrumental maps are generated using the filter-wheel-closed data \cite{Zhang25} scaled by the observed 8--11\kev\ count rate, which is dominated by particle background.
We generate exposure maps using {\it fxtexpogen} and mask bad pixels and edge pixels with {\tt RAWY} $< 20$, where the photon collection is influenced by the shade of storage area. The exposure maps are weighted by the effective area to correct for the differences in response between two the FXT modules. 
We use the CIAO script {\it dmimgthresh} to exclude regions with low effective exposure time (less than 15\% of the maximum). The counts maps, exposure maps, and instrument background maps of different observations are stacked by {\it dmimgfilt}. 

One needs to mask point-like sources to minimize their influence to the detection of faint diffuse emission.  
We first employ the {\it ciao} script {\it wavdetect} for source detection.
Based on the default 3$\sigma$ source regions produced by {\it wavdetect}, we examined each of them and enlarge the region where necessary to prevent bright wing emission leaking into the field. 
Then, we use the {\it ciao} script {\it vtpdetect}, which is a non-parametric tool to detect photon clustering in the image. 
This approach allows for a more complete detection of sources with irregular morphologies, e.g., sources in close proximity to each other or near the edge of the field of view. 
These procedures ensure a conservative mask, thereby effectively eliminating contamination from point-like sources (see also \cite{Schellenberger15}). Despite some faint residuals after masking, their contribution is negligible compared to the source regions with large angular sizes for spectral analysis.
The masked regions (see Supplementary Figure~\ref{fig:pls}) are then re-filled using {\it ciao} commands {\it roi} and {\it dmfilth}. After the subtraction of background and point-like source, and correction for the vignetting effect, we eventually obtain the flux images of diffuse emission, which are then adaptively smoothed with {\it dmimgadapt}.

\begin{figure}
    \centering    \includegraphics[width=\textwidth]{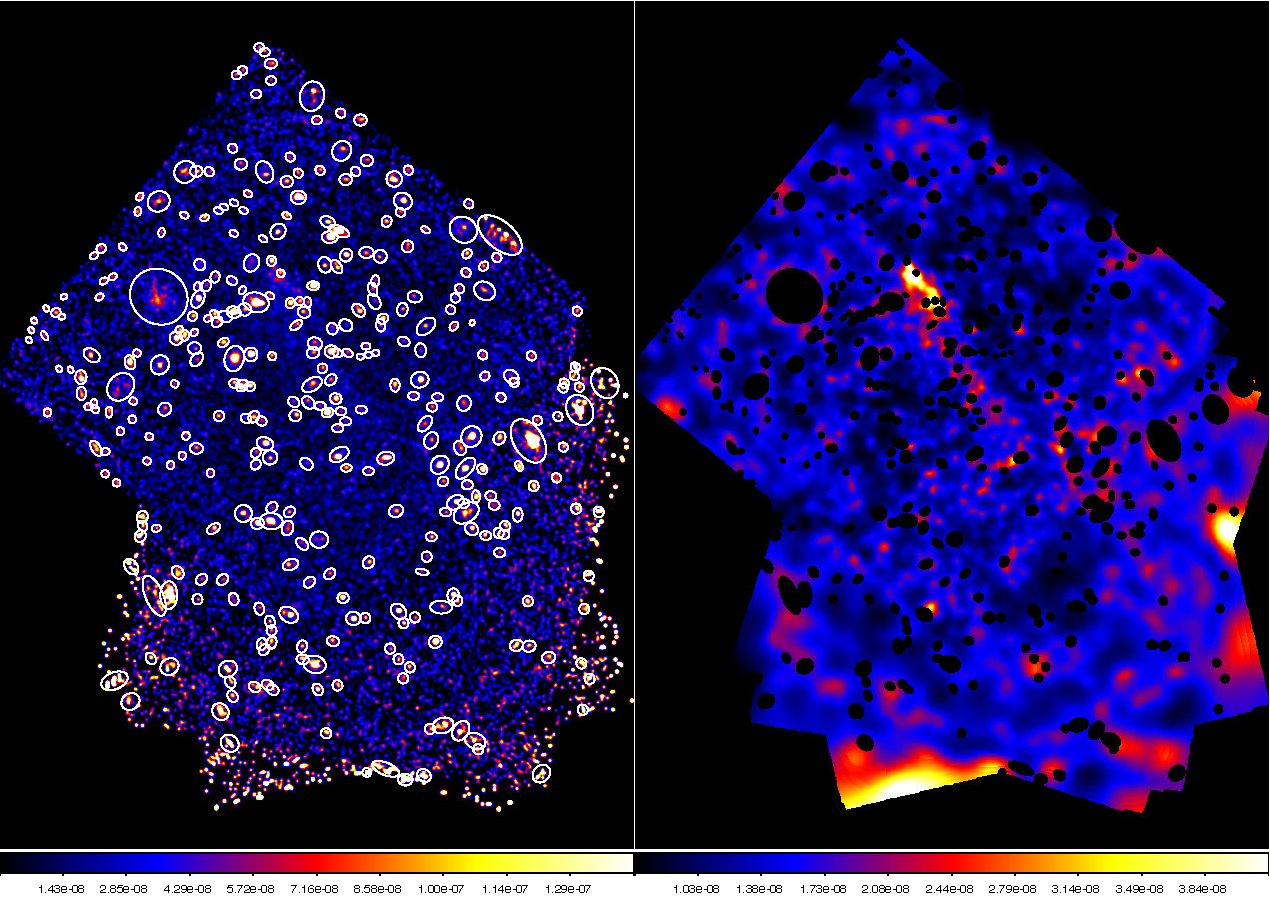}
    \caption{Photon flux image in units of counts\,s$^{-1}$\,cm$^{-2}$ overlaid with the regions for source masking (left) and adaptively smoothed image with source masked (right). Both images are particle background subtracted and vignetting corrected. Bright features near the field of view edge  result from large fluctuations caused by low effective exposure and large PSF, and have negligible impact to our analysis. }
    \label{fig:pls}
\end{figure}

We select the 1--5\,keV band because it optimally highlights the potential diffuse non-thermal signals. Below $\sim1$\,keV, the observed X-ray photons are mainly from the background diffuse hot gas, while above $\sim5$\,keV, the particle background dominates. 
Besides, the effective area of FXT peaks near 1--2\,keV. Therefore, the 1--5\kev\ band well balances the spectral characteristics and instrument performances. 

\subsection{Spectral analysis}

For spectral extraction, an even larger circular mask (15$^{\prime\prime}$ in radius) is used for point-like sources. 
We extract the X-ray spectra and response files from the source region and  a nearby sky background region, using the commands {\it xselect}, {\it fxtrmfgen} and {\it fxtarfgen}.  
Spectra from individual observations are grouped to a minimum of 50 counts per bin with {\it ftgrouppha} to ensure the use of chi-square statistics then jointly fitted using Xspec (ver.\ 12.14.0b \cite{Arnaud96}).

\section{XMM-Newton/MOS}

Deep XMM-Newton \cite{Jasen01} observations have been performed to study the pulsar PSR~J1740+1000 and its PWN, but covered only the northeastern portion of  1LHAASO J1740+0948u.
We retrieve the MOS \cite{turner01} observations which are in the Full Frame mode; the pn data \cite{Struder01}  in the small window mode are not used.
The MOS data (see Supplementary Table~\ref{tab: xmmobs}) are reprocessed by XMM-Newton SAS (ver.\ 20 \cite{Gabriel04}) along with the XMM-Newton Extending Source Analysis Software (ESAS). The original data are reprocessed with {\it emchain} and high background intervals were excluded with {\it mos-filter}. The tasks {\it mos-spectra} and {\it mos\_back} are used to produce images and spectra with instrument background subtracted. Images of different observations and cameras are stacked with {\it merge\_com\_xmm} and then adaptively smoothed with {\it adapt\_merge}. Due to the loss of CCDs 3 and 6 in MOS1, only MOS2 data are used for spectral analysis. 
XMM-Newton data can better constrain the PWN component closer to the pulsar, thereby offering an independent verification of the FXT results. Also, the slightly higher angular resolution of XMM-Newton MOS ($15''$) and deep exposure enable to better identify and subtract point-like sources. 
The spectra of different observations are fitted jointly.

\begin{table}
    \centering
    \caption{XMM-Newton/MOS observations of PSR~J1740+1000. The exposure time is flare-filtered.}
    \begin{tabular}{cccc}
    \hline
        Obs. ID & Obs. Date & MOS1 Exposure & MOS2 Exposure \\ 
        & & (ks) & (ks) \\
        \hline
        0403570101 & 2006-09-28 & 39.14 & 39.73 \\
        0403570201 & 2006-09-30 & 24.90 & 24.93 \\
        0803080201 & 2017-09-20 & 90.20 & 92.46 \\
        0803080301 & 2017-10-04 & 102.25 & 105.32 \\
        0803080401 & 2018-03-05  & 67.19 & 72.53 \\
        0803080501 & 2018-04-04 & 111.61 & 119.48 \\ \hline
        Total & ~ & 435.29 & 454.45 \\ \hline
    \end{tabular}
    \label{tab: xmmobs}
\end{table}

\section{X-ray background}

As all three regions span several arcminutes, the vignetting effect and particle background can no longer be regarded as constant within a single region, let alone between separated regions (i.e., between the source and sky background regions). The particle background is not affected by the vignetting, whereas the sky background is. Consequently, a direct subtraction of background spectra \cite{Kargaltsev08, Brunelli25, Gagnon25} might be problematic for very extended, faint sources. To correctly handle the background and response, we adopted a double-background subtraction approach (instrumental + sky). After subtracting the instrumental quiescent particle background (QPB) in each region, we modeled the sky background and then added this component when fitting the source spectra. The particle background spectra of XMM-Newton were generated by {\it mos\_back}, while for FXT, we scaled the filter-wheel-closed data using the observed 8--11 keV data, where the instrumental background dominates \cite{Zhang25}. Besides, we only focused on the 1--7\,keV band in the spectral fitting of source regions, where the synchrotron emission dominates, to reduce the systematic uncertainties due to the high background.

We first modeled the sky background --- cosmic X-ray background and Galactic diffuse emission --- using XMM-Newton data, given the deep exposure and better statistics. 
We selected a background region near the optical axis of XMM-Newton, next to region A. The sky background is fitted in 0.4--7\,keV and consists of three components, (1) an unabsorbed low-temperature ($kT\sim$0.1\,keV) local hot gas (e.g., the local hot bubble) with variable temperature and normalization ({\it apec}), (2) a Galactic hot gas (e.g., the Galactic halo) with variable temperature, abundances, and normalization subject to Galactic absorption ({\it TBabs}) with a column density fixed at $5\times10^{20}$\psqcm\ \cite{Kargaltsev08}, and (3) a non-thermal power-law component with the same amount of Galactic absorption to mimic the cosmic X-ray background (CXB) with a fixed photon index of 1.46 \footnote{according to the XMM-ESAS cookbook https://heasarc.gsfc.nasa.gov/docs/xmm/esas/cookbook/xmm-esas.html} and a variable normalization. Besides, the instrumental fluorescence Al-K and Si-K line of MOS are reproduced by two Gaussian components around 1.49 and 1.76\,keV, respectively. It should be noted that a good mimic of the sky background is sufficient so that a detailed discussion of the sky model parameters is not necessary. 
The background fitting gives a reduced chi-square of $\chi^2_r=1.1$ with ${\rm dof} = 443$. 

We extracted an FXT background spectrum from a region with a radius of $20'$, excluding any source regions and point-like sources. The region avoids the boundary areas of the CCD, where the background level has larger fluctuations due to the worse spatial resolution and lower effective area. 
The FXT background in the 1--7 keV band can be well reproduced by the XMM-Newton sky background model, which is adopted in the subsequent fitting.

For the source regions, the XMM-Newton sky background model is scaled according to the {\it backscal} keyword, which corresponds to the sky area of each region. The parameters of the sky background are fixed, and an additional absorbed power-law component is added to account for the synchrotron emission. The absorption column density is fixed at $5\times10^{20}$\,cm$^{-2}$ \cite{Kargaltsev08}, and the effect of absorption is negligible above $\sim1$ keV. To facilitate comparison with previous studies, we used the unabsorbed flux in the 0.5--8 keV band as a measure of the intensity. These, along with the photon index, are the only free parameters in the fits. In addition to the measured flux, we also calculated the flux density of each region and apply corrections for flux loss as a result of masking point-like sources.

\section{LHAASO}

\subsection{Observations and data reduction}

LHAASO is a dual-purpose complex of particle detectors designed for studying cosmic rays (CRs) and $\gamma$-rays \cite{2010ChPhC..34..249C}. Its $\gamma$-ray detection system comprises two sub-arrays, the Square Kilometer Array (KM2A) and the Water Cherenkov Detector Array (WCDA), which collectively span an energy detection range from sub-TeV to 10 PeV ($10^{16}$ eV). Among these, KM2A stands out as the most sensitive instrument for detecting UHE $\gamma$-ray emission, leading the UHE $\gamma$-ray era with detections of $\sim40$ UHE sources \cite{Cao24}.

In this work, we utilized over 1944 days of data from LHAASO-KM2A and over 1484 days of data from LHAASO-WCDA. Adhering to the LHAASO analysis pipeline, the data preprocessing procedures, including event selection, reconstruction, and $\gamma$-proton discrimination, followed the methodologies outlined in previous performance evaluation publications by LHAASO \cite{2021ChPhC..45b5002A,2021ChPhC..45h5002A,2025APh...16403029C}. We selected the $\gamma$-like events (after CR background rejection) within a $5^\circ \times 5^\circ$ region centered on 1LHAASO J1740+0948u. The data were binned to $0.1^\circ \times 0.1^\circ$ spatial bins across 17 energy bins, covering an energy range from 10 TeV to above 1 PeV. Furthermore, we estimated the CR background in each bin using the integral method \cite{2004ApJ...603..355F}. For the analysis, we employed the $\gamma$-tool developed by the LHAASO collaboration to perform the binned likelihood analysis. As the source is located at a high Galactic latitude ($b = 20.3^\circ$), we ignored the Galactic diffuse emission.

\subsection{Morphology analysis}

A 3-dimensional (3D) likelihood algorithm was used to fit the morphology and spectrum of the source simultaneously, and the test statistic (TS) was used to evaluate the detection significance of the source, defined as the ratio of the maximum likelihood,
$ {\rm TS} = 2 \ln (\mathcal{L} (\theta_1)/ \mathcal{L}(\theta_0))$,
where $\theta_0$ and $\theta_1$ represent the parameters of the null and the alternative hypotheses, respectively. Based on Wilks' Theorem, the TS follows a $\chi^{2}_{n}$ distribution, where $n$ is the degree of freedom (dof) derived from the difference in the number of free parameters between the models.  We employed the difference in source significance ($\Delta$TS) to compare the spatial and/or spectral models of the source. It is important to note that the difference in TS cannot be used to quantitatively determine the preferred model when the models are not nested. Alternatively, the Akaike information criterion test (AIC \cite{1974ITAC...19..716A}) can be considered. The AIC is defined as AIC = $2k - 2 ln\mathcal{L}$, where $k$ is the number of parameters in the model. In this context, the best hypothesis is considered to be the one that minimizes the AIC. A qualitative strength of evidence rule to assess the significance of a model is based on the difference in AIC ($\Delta$AIC) between the two models. If $\Delta\rm\  AIC > 5$, then it is considered strong evidence against the model with a higher AIC and 
$\Delta\rm\ AIC > 10$ constitutes decisive evidence \cite{2007MNRAS.377L..74L}.

To fit the $\gamma$-ray emission morphology, we explored various geometrical models, including a point source, a Gaussian, and an elliptical Gaussian, to characterize the LHAASO emission.  To investigate the possible energy-dependent morphological variation, we employ a double-elliptical Gaussian model, where the spatial parameters are independently fitted for the energy bands of 10 -- 100 TeV and above 100 TeV, respectively.
Additionally, we incorporated a morphology template obtained from EP observations (flux densities taken from the three regions) to test for spatial correlations between $\gamma$-ray and X-ray.  

\begin{table}[ht]
    \centering
    \caption{LHAASO morphology analysis above 10 TeV.}
    \label{tab:lhaasobs}
    \renewcommand{\arraystretch}{1.2}
\begin{tabular}{lcccccccc}
    \hline
        Model & R.A. & Decl. & $95\%$ Position & Semi-major & Semi-minor & Rotation & TS &  AIC \\
        ~ & (J2000) & (J2000) & Error (deg) & Axis (deg) & Axis (deg) & Angle (deg) & ~ & ~ \\ \hline
\hline
Point Source & 265.04 & 9.82 & 0.04 & $\cdots$ & $\cdots$& $\cdots$ & 494.8 & $\cdots$ \\
Gaussian & 265.03 & 9.81 & 0.04 & 0.11$\pm$0.02 & $\cdots$& $\cdots$& 500.1 & -3.3 \\
Elliptical Gaussian & 265.03 & 9.81 & 0.05 & 0.18$\pm$0.03 & $<$0.12 & 49.6$\pm$8.8 & 515.6 & -14.8 \\
\hline
Double-Elliptical & 265.01 & 9.81 & 0.05 & 0.20$\pm$0.04 & $<$0.15 & 58.2$\pm$10.6 &520.8 & -10.0\\
     Gaussian & 265.07 & 9.80 & 0.07 & 0.17$\pm$0.04 & $<$0.10 & 36.7$\pm$14.1 & & \\
\hline
X-ray Template & $\cdots$ & $\cdots$ & $\cdots$ & $\cdots$  & $\cdots$ & $\cdots$& 507.3 & -16.5 \\
\hline
\hline
\end{tabular}
\begin{tablenotes}
\item Note: $P_{95}$ is the statistical positional uncertainty at a 95\% confidence level. The semi-major axis corresponds to an encircled 39\% flux region for Gaussian model. The upper limit of the extension along the semi-minor axis corresponds to 95\% confidence level. The AIC values are subtracted from that of the point model for a clear comparison. The rotation angle are from north to east. Double-elliptical Gaussian model, where the spatial parameters are independently fitted for 10--100\,TeV and above 100\,TeV. respectively.
\end{tablenotes}
\end{table}

The parameters of the morphology tests are summarized in Supplementary Table~\ref{tab:lhaasobs}.
The Gaussian model demonstrates a statistically significant improvement over the point-source model, with a TS enhancement of 5.3 (corresponding to $2.3 \sigma$ significance). This improvement is further enhanced with the elliptical Gaussian model, which achieves an extended significance of $3.9 \sigma$ ($\Delta$TS = 20.8), indicating an extended aspherical  morphology. 
Compared to the elliptical Gaussian model, the improvement in TS with the Double-elliptical Gaussian model is negligible, implying that significant energy-dependent morphologies cannot be resolved with current observations. 
Given the non-nested nature of the X-ray tail model with respect  to other models, we perform model selection the AIC. The difference in AIC ($\Delta$AIC = 16.5) decisively favors the X-ray tail template over the point-source model, providing strong evidence against the latter and confirming the extended morphology of $\gamma$-ray emission. The anisotropic characteristics of the $\gamma$-ray emission is further supported by $\Delta$AIC = 13.2 between the X-ray template model and the standard Gaussian model. While the X-ray template model exhibits an AIC difference of 1.7 over the elliptical Gaussian model, the latter remains statistically plausible and cannot be definitively excluded. 

\begin{figure}[ht]
    \centering
    \includegraphics[width=\textwidth]{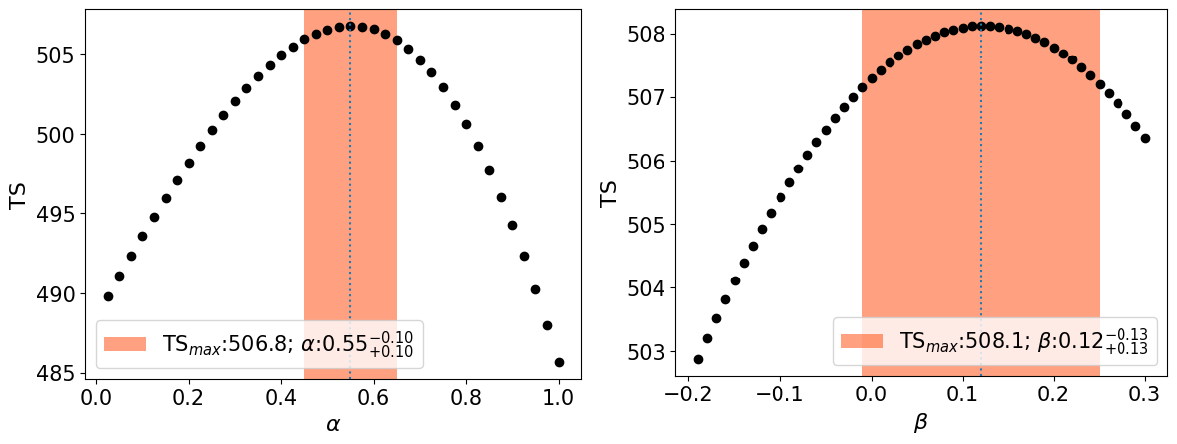}
    \caption{Left: the likelihood profile for the power-law spatial models. Right: the likelihood profile for the scaled X-ray profile spatial models.}
    \label{fig:spatial_index}
\end{figure}

Theoretically, $\gamma$-ray morphology should not perfectly align with the X-ray profile due to potential variations in magnetic field conditions. To test this scenario, we modeled a series of spatial templates assuming $\gamma$-ray surface brightness follows a power-law distribution $S \propto r^{-\alpha}$, where $r$ is the distance to the pulsar and $\alpha$ is the index of the power-law spatial model. The $\gamma$-ray emission region is assumed to be the same as the X-ray extension region, i.e., a width of $6'$ and a length of $32'$. We generated templates with indices $\alpha$ ranging from 0 to 1.0 in steps of 0.025. After convolution with the LHAASO PSF, these templates were evaluated within a 3D maximum-likelihood framework to compute likelihood profiles for the value $\alpha$, allowing the derivation of parameter constraints. As shown in the Supplementary Fig~\ref{fig:spatial_index}, the parameter can be well constrained to $\alpha=0.55\pm0.10$. Alternatively, we considered scaled X-ray templates to model $\gamma$-ray morphology. We assumed a $\gamma$-ray intensity profile scaling with the X-ray profile according to $S_\gamma\propto r^{-\beta} S_x$,  where $r$ is the distance to the pulsar and $\beta$ is the index of the power-law shaped scale factor. The index $\beta$ was sampled from $-0.2$ to 0.3 in steps of 0.01. Each template was convolved with the LHAASO PSF and evaluated within our likelihood framework to generate corresponding likelihood profiles. This analysis constrained the scaling parameter to  $\beta = 0.12 \pm 0.13$.

\subsection{Spectral analysis}

For the spectral analysis of 1LHAASO J1740+0948u, we conducted a maximum likelihood fitting in the energy range above 10 TeV, employing the X-ray tail model described previously. The spectral points were derived through a maximum likelihood analysis performed in each energy bins. For the bins with a TS value below 4 (2$\sigma$), we provide the upper limits on the flux at a 95\% confidence level. 
We compared three spectral models for 1LHAASO J1740+0948u: a power-law (PL), a power-law with exponential cutoff (PLC), and a log-parabola (LP). As shown in Supplementary Table~\ref{tab:lhaasoSED}, introducing spectral curvature significantly improved the fit ($>8.9\sigma$). However, the LP and PLC models could not be distinguished due to their small AIC difference ($\Delta {\rm AIC} = 0.2$).  Tentatively, we adopted the LP model, consistent with the previous study \cite{Cao2025_1740}, and estimated an integral energy flux above 10\,TeV as $F({> 10\rm\ TeV}) \approx  5.5 \times 10^{-13 }\rm\ erg\ cm^{-2}\ s^{-1}$. 

\begin{table}[ht]
  \centering
 \caption{Spectral Fitting Results for 1LHAASO J1740+0948u above 10 TeV}

 \label{tab:lhaasoSED}
 \begin{tabular}{lcccccccc}
 \hline
\textbf{Model}&
\textbf{$N_0$} &
\textbf{$\alpha$}&
\textbf{$\beta$}&
\textbf{$E_{0}$}&
\textbf{$E_{c}$}&
\textbf{TS} &
\textbf{AIC}\\
~ &($10^{-16}$\,cm$^{-2}$\,s$^{-1}$\,TeV$^{-1}$) & & &(TeV) &(TeV) & & \\

\hline
PL & 2.98$\pm$0.21 & 2.48$\pm$0.04 & $\cdots$ & 20.0 & $\cdots$ & 426.7 & $\cdots$ \\
LP & 1.87$\pm$0.15 & 2.15$\pm$0.16 & 1.25$\pm$0.27 & 30.0 & $\cdots$ & 507.3 & $-$78.6 \\
PLC & 2.99$\pm$0.52 & 1.46$\pm$0.23 & $\cdots$ & 30.0 & 50.8$\pm$10.7 & 507.5 & $-$78.8 \\
\hline
 \end{tabular}  
\begin{tablenotes}
\item Note: PL stands for the power-law model defined by $dN/dE=N_0 (E/E_0)^{- \alpha}$, LP represents the log-parabola model defined by $F(E)=N_0 (E/E_0)^{-(\alpha+\beta \log_{10}(E/E_0))}$, and PLC represents the power-law with exponential cutoff defined by $dN/dE=N_0 \times (E/E_0)^{- \alpha} \exp[-(E/E_c)]$. For the LP model, the latter value listed in $\alpha$ column is the $\beta$ parameter. The AIC value is subtracted from that of the PL spectrum for a clear comparison.
\end{tablenotes}
\end{table}

\subsection{Systematic uncertainties}
To assess the robustness of our results, we performed systematic checks following the methodology of Ref.~\cite{Cao24}. The dominant systematic uncertainty in position measurement originated from pointing errors, estimated at $0.04^\circ$ for both WCDA and KM2A. Additional PSF-related systematic uncertainties were approximately $0.05^\circ$ for WCDA and $0.08^\circ$ for KM2A. 

For extension measurements, PSF uncertainties dominate the systematic error, propagating to the measured extension as:
\[
\sigma_{\mathrm{ext,sys}} = 0.01^\circ,
\]
calculated using:
\[
\sigma_{\mathrm{ext,sys}} = \sqrt{\mathrm{ext}^2 + \sigma_{\phi}^2} - \mathrm{ext},
\]
where $\mathrm{ext}$ denotes the intrinsic source extension and $\sigma_{\phi}$ represents the systematic uncertainty of PSF.

Flux systematic uncertainties of 8\% were attributed to atmospheric modeling in Monte Carlo simulations. We further evaluated Galactic diffuse emission contributions by incorporating a dust template, finding a negligible flux impact of $<2\%$.

\setlength{\bibsep}{1pt}
\small
\bibliographystyle{scpma}
\bibliography{J1740}